\documentclass[reprint,pra,longbibliography]{revtex4-1}
\usepackage{graphicx}
\usepackage{color}
\usepackage{bm}
\usepackage{amsmath}
\usepackage{amssymb}

\begin{document}

\title{Resonant field enhancement near bound states in the continuum
  on periodic structures}

\author{Zhen Hu}
\affiliation{Department of Mathematics, Hohai University, Nanjing,
  Jiangsu, China}

\author{Lijun Yuan}
\affiliation{College of Mathematics and Statistics, Chongqing Technology and Business University, Chongqing, 
China}

\author{Ya Yan Lu}
\affiliation{Department of Mathematics, City University of Hong Kong, 
  Kowloon, Hong Kong, China} 
\date{\today}

\begin{abstract}
On periodic structures sandwiched between two homogeneous media,
a bound state in the continuum (BIC)  is a guided Bloch mode with a
frequency  within the radiation continuum. BICs are useful, since they give rise
to high quality-factor ($Q$-factor) resonances that enhance
local fields  for diffraction problems with given incident waves. For 
any BIC on a periodic structure, there is always a surrounding family of
resonant modes with $Q$-factors approaching infinity. 
We analyze field enhancement around BICs using analytic and numerical methods. 
Based on a perturbation method, we show that field enhancement is
proportional to the square-root of the $Q$-factor, and it depends
on the adjoint resonant mode and its coupling efficiency with incident
waves. Numerical results are presented to show different
asymptotic relations between the field enhancement and the Bloch
wavevector for different BICs. Our study provides a useful guideline for
applications relying on resonant enhancement of local fields. 
\end{abstract}

\maketitle 

\section{Introduction}
Bound states in the continuum (BICs) for photonic systems  are
attracting significant research interest mainly because they lead to
resonances of extremely high  
quality factors ($Q$-factors) \cite{hsu16,kosh19}. A BIC  is a trapped
or guided mode with 
a frequency in the radiation continuum where radiative waves can
propagate to or from infinity \cite{neumann29}, and it can only exist
on a lossless 
structure that is infinite in at least one spatial direction. 
Many recent works are concerned with BICs on periodic
structures (with one or two periodic directions) sandwiched between
two homogeneous media
\cite{bonnet94,padd00,tikh02,shipman03,port05,shipman07,mari08,lee12,hsu13_2,bulg14b,yang14,zhen14,hu15,gao16,gan16,li16,ni16,yuan17,bulg17pra,hu18,doel18,zhang18}. 
On such a periodic structure,  a BIC  is a Bloch
mode that decays exponentially in the surrounding homogeneous media,
but unlike ordinary  guided modes below the lightline, it co-exists
with plane waves above the lightline. These plane waves have the same
frequency and wavevector as the BIC, and can propagate to or from
infinity in the surrounding media. 
 Importantly, a BIC can be regarded as a resonant
mode with an infinite $Q$-factor \cite{hsu13_2}.  When the structure is
perturbed slightly, a 
BIC usually (but not always) becomes a resonant mode with a large
$Q$-factor \cite{yuan17_4,kosh18,lijun19pr}. On periodic structures, a BIC is surrounded by resonant
modes that depend on the Bloch vector continuously. The $Q$-factors of
these resonant modes tend to infinity as the Bloch wavevector
approaches  that of the BIC \cite{hsu13_2}. The relation  between the
$Q$-factor and the Bloch wavevector has been analyzed in a number of
papers \cite{lijun17pra,lijun18pra,jin19,lijun19pr}.

Strong local fields induced by high-$Q$ resonances
are essential to applications such as lasing \cite{kodi17} and sensing
\cite{romano19,yesi19}, and can be used enhance emissive processes and
and nonlinear effects \cite{lijun16pra,lijun17pra,lijun19siam}. For 
lossless dielectric structures, the 
$Q$-factor (denoted by $Q$) of a resonant mode accounts for radiation losses 
only, and the field enhancement, defined as the ratio of the
maximum amplitudes of the total and incident waves, is known to be proportional to
$\sqrt{Q}$. Therefore, using the asymptotic relations between the
$Q$-factors and the wavevectors, the field enhancement caused by
a high-$Q$ resonance near a BIC is known qualitatively \cite{shipman05,mocella15,yoon15}.
 In this paper, we use a perturbation method to derive a rigorous formula for field
enhancement and perform a detailed numerical study for field
enhancement around a few BICs with distinct asymptotic behavior. 
The perturbation theory is developed for two-dimensional (2D) 
periodic structures with one periodic direction (i.e., 1D
periodicity). The formula reveals not 
only the square-root dependence on the $Q$-factor, but also the
relevance of adjoint resonant modes and their coupling efficiency
with incident waves. The numerical examples are carried out for a
few BICs for which the corresponding $Q$-factors 
have different asymptotic behaviors. 

The rest of this paper is organized as follows. In Sec.~II, we briefly
recall the definitions and properties of BICs and resonant modes on
2D periodic structures, and establish a useful formula for the
$Q$-factor. In Sec.~III, we derive a formula for field enhancement
using a perturbation method. Numerical examples for four BICs with
very different properties are presented in Sec.~IV. The paper is
concluded with a brief discussion in Sec.~V. 


\section{Resonant modes around BICs}

Many recent studies on photonic BICs are concerned with 
dielectric periodic structures. Two-dimensional periodic structures
that are uniform in 
one spatial direction and periodic in another direction are
relatively simple to analyze theoretically, but they still capture the
nontrivial physics involving the BICs  
\cite{bonnet94,shipman03,port05,shipman07,mari08,bulg14b,hu15,yuan17,hu18}. 
 We consider
2D periodic structures that are invariant in $z$, periodic in $y$ with 
period $L$, and sandwiched between two homogeneous media given in
$x<-D$ and $x>D$ for a positive $D$, respectively. Let ${\bf  r} =
(x,y)$ and $\epsilon = \epsilon({\bm r})$ be the 
dielectric function for such a periodic structure and the
surrounding media, then $\epsilon$ is real and positive, and 
\begin{equation}
  \label{periodic}
  \epsilon(x,y+L) = \epsilon({\bm r}) 
\end{equation}
for all ${\bm r}$. For simplicity, we assume the surrounding medium is
vacuum, thus 
\begin{equation}
  \label{air}
\epsilon({\bm r}) = 1, \quad \mbox{if}\ |x|>D.
\end{equation}
For the $E$-polarization and a time harmonic field with the time
dependence $e^{-i \omega t}$ ($\omega$ is the angular frequency), the
$z$ component of the electric field, 
denoted as $E_z$ or $u$, satisfies the 
following 2D Helmholtz equation
\begin{equation}
  \label{helm2d}
\partial_x^2 u + \partial_y^2 u + k^2 \epsilon({\bm r}) u = 0,
\end{equation}
where $k = \omega/c$ is the freespace wavenumber and $c$ is the speed
of light in vacuum.  

A BIC on the periodic structure is a solution of Eq.~(\ref{helm2d})
for a real $\omega > 0$, given in the form of a Bloch mode
\begin{equation}
  \label{bloch}
  u({\bm r}) = \phi({\bm r}) e^{ i \beta y},
\end{equation}
where $\phi$ is periodic in $y$ with period $L$, $\phi \to 0$
exponentially as $|x| \to \infty$,  $\beta$ is the real Bloch
wavenumber, and $k > |\beta|$. Due to the periodicity in $y$, $\beta$
can be restricted by $|\beta| \le \pi/L$. If $\beta=0$, the BIC is a
standing wave, otherwise, it is a propagating BIC. Since the
lightlines (in the $\beta$-$k$ plane) are defined as $k=\pm \beta$, a
BIC is a guided mode above the 
lightline. For $|x|>D$,  any Bloch mode given by Eq.~(\ref{bloch}) 
can be expanded in plane waves as 
\begin{equation}
  \label{pwaves}
u({\bm r}) = \sum_{m=-\infty}^\infty c_m^\pm e^{i [ \beta_m y \pm
  \alpha_{m} (x\mp D)]}, \quad \pm x > D, 
\end{equation}
where $\beta_0=\beta$, and 
\begin{equation}
  \label{albeta}
\beta_m = \beta + \frac{2\pi m}{L}, \quad \alpha_{m} = \sqrt{k^2 -
  \beta_m^2}. 
\end{equation}
Most (but not all) BICs are found for $k$ satisfying 
\begin{equation}
  \label{1chanel}
|\beta| < k  < \frac{2\pi}{L}-|\beta|. 
\end{equation}
In that case, all $\alpha_m$ for $ m \ne 0$ are pure imaginary with 
positive imaginary parts, and only $\alpha_0$ is real. Since a BIC
must decay exponentially as $|x| \to \infty$, if condition (\ref{1chanel})
is satisfied, then we must have $c_0^+ = c_0^-  = 0$. 

A resonant mode (also called resonant state, quasi-normal mode, guided
resonance, or scattering resonance) on
the periodic structure is a 
solution of Eq.~(\ref{helm2d}) that radiates  
power outwards as $|x| \to \infty$ \cite{fan02,amgad19}. 
 Since $\epsilon$ is real and
energy is conserved, the frequency $\omega$ of a resonant mode must
have a negative imaginary 
part, so that it can decay with time as it radiates power
to infinity. 
The $Q$-factor of a resonant mode can be defined as 
$Q = -0.5 \mbox{Re}(\omega)/\mbox{Im}(\omega)$. Expansion 
(\ref{pwaves}) is still valid, provided that all $\alpha_m = \sqrt{
  k^2 - \beta_m^2}$ are  
properly defined to maintain continuity as $\mbox{Im}(\omega)$ tends
to zero. This can be achieved by choosing the 
negative imaginary axis (instead of the negative real axis) as the
branch cut for the complex square root. More precisely, if 
$\eta = |\eta| e^{ i \theta}$ for $-\pi/2 < \theta \le 3\pi/2$
(instead of $-\pi < \theta \le \pi$), then $\sqrt{\eta}  =
\sqrt{|\eta| }  e^{ i \theta /2}$. If $\mbox{Re}(k)$ satisfies
condition (\ref{1chanel}) and $\mbox{Im}(k)$ is small, then all $\alpha_m$
for $m \ne 0$ have positive imaginary parts, and $\alpha_0$ has
a positive real part and a small negative imaginary part. In that
case, the plane wave $\exp[ i (\beta y + \alpha_0 x)]$ radiates power in
the positive $x$ direction and blows up as $x \to +\infty$. The
coefficients $c_0^{\pm}$ of a resonant mode  should be nonzero. 
The resonant modes form bands with complex frequency $\omega$
depending on real Bloch wavenumber $\beta$.  A BIC corresponds to a special
point on the dispersion curve of a band of resonant modes, where
$\omega$ becomes real. Therefore, a BIC can be regarded as special
resonant mode with an infinite $Q$-factor. 

Let $\omega_\circ$ and $\beta_\circ$ be the frequency and Bloch wavenumber of
a BIC respectively, and $\omega$ be the complex frequency of a
resonant mode near the BIC for a $\beta$ near
$\beta_\circ$. Perturbation theories provide approximate formulas for
$\omega$ and the $Q$-factor assuming $|\beta-\beta_\circ|$ is
small. In general, we have 
\begin{eqnarray}
\label{greal1}
 \mbox{Re}(\omega) - \omega_\circ  \sim  \beta-\beta_\circ,  \\
\label{gimag2}
  \mbox{Im}(\omega)    \sim  (\beta-\beta_\circ)^2, \\
\label{gqfactor2}
 Q  \sim   1/|\beta - \beta_\circ|^2. 
\end{eqnarray}
Special results have been derived for standing waves on periodic
structures with a reflection symmetry in the periodic
direction \cite{lijun17pra,lijun18pra,lijun19pr}. 
 Assuming the periodic structure is symmetric in $y$  
(i.e., $\epsilon$ is even in $y$), 
a standing wave  is either symmetric in $y$
(even in $y$) or antisymmetric in $y$ (odd in $y$). For both cases, it
is known that 
\begin{equation}
  \label{sreal2}
  \mbox{Re}(\omega) - \omega_\circ \sim  \beta^2.  
\end{equation}
Moreover, for a symmetric standing wave, we have 
\begin{equation}
  \label{simag4}
  \mbox{Im}(\omega)  \sim  \beta^4.
\end{equation}
For a typical antisymmetric standing wave, 
Eq.~(\ref{gimag2}), i.e.,  $\mbox{Im}(\omega)\sim \beta^2$, 
 is valid, but under special conditions,
$\mbox{Im}(\omega)$ satisfies 
\begin{equation}
  \label{simag6}
\mbox{Im}(\omega)  \sim \beta^6.  
\end{equation}
Therefore, depending on the nature of the standing wave, the
$Q$-factor follows different scaling laws, i.e.,  $1/\beta^2$,
$1/\beta^4$, or $1/\beta^6$.  

The $Q$-factor of a resonant mode is often defined as
the ratio between the energy stored in a cavity and the power loss,
multiplied by the resonant frequency (real part of the complex
frequency). 
For our 2D periodic structure, the
cavity can be chosen as the rectangle
\begin{equation}
  \label{defOmg}
\Omega = \{ (x,y) \, : \, |x| < D, \, |y| < L/2 \}.  
\end{equation}
In Appendix A, we derive a formula for the $Q$-factor, 
i.e., Eq.~(\ref{ourQ}) below,  based on a 
wave-field splitting outside the cavity. If  there is only one radiation
channel, i.e., $\alpha_0$ is in the 
fourth quadrant close to the positive real axis, and all $\alpha_m$ for
$m\ne 0$ are in the second quadrant close to the positive imaginary
axis, then, the wave field outside the cavity can be written as
\begin{equation}
  \label{split}
  u({\bm r}) = u_p ({\bm r}) + u_e ({\bm r}), \quad |x| > D, 
\end{equation}
where $u_p$ is the term for $m=0$ in the right hand side of
Eq.~(\ref{pwaves})  and  
$u_e$ is the sum of all other terms for $m\ne 0$. In that case, the
$Q$-factor satisfies 
\begin{equation}
  \label{ourQ}
\frac{1}{Q} = 
\frac{ L \mbox{Re}(\alpha_0)}{ [ \mbox{Re}(k) ]^2} \cdot 
\frac{ |c_0^+|^2 + |c_0^-|^2} 
{\int_\Omega \epsilon |u|^2 d{\bm r}  + 
\int_{\Omega_e}  |u_e|^2 d{\bm r}},
\end{equation}
where $\Omega_e$ is the union of two semi-infinite strips given by
$|x|>D$ and $|y| < L/2$.  The first and second integrals in the
denominator are proportional to the electric energy stored in the
cavity and the electric energy of the evanescent field $u_e$ outside the
cavity. The numerator is proportional to the power radiated out by the
plane wave $u_p$. Assuming the resonant mode is scaled such that 
\begin{equation}
  \label{max1}
\max_{ {\bm r} \in \Omega} |u({\bm r})| = 1,  
\end{equation}
then $c_0^+$ and $c_0^-$ are dimensionless quantities, and
Eq.~(\ref{ourQ}) gives rise to  
\begin{equation}
  \label{c0pm} 
|c_0^+|^2 + |c_0^-|^2 \sim \frac{1}{Q}.  
\end{equation}

By reciprocity, corresponding to one resonant mode $u$ with a real
Bloch wavenumber $\beta$ and a complex frequency $\omega$, there is
another resonant mode (the adjoint resonant mode) $v$ with Bloch
wavenumber $-\beta$ and the same 
complex frequency. We write $v$ as 
\begin{equation}
  \label{recimode1}
v({\bm r})  = \psi({\bm r}) e^{-i \beta y}, 
\end{equation}
and expand $v$ as 
\begin{equation}
  \label{vpwave}
v({\bm r}) = \sum_{m=-\infty}^\infty d_m^{\pm} e^{i [ - \beta_m y \pm 
  \alpha_{m} (x\mp D)]}, \quad \pm x > D.  
\end{equation}
Apparently, Eq.~(\ref{ourQ}) remains valid when $u$, $u_e$, $c_0^\pm$
are replaced by $v$, $v_e$ (similarly defined as $u_e$) and
$d_0^\pm$. If we scale $v$ such that $\max_{(x,y) \in \Omega} |
v(x,y)| = 1$, then 
\begin{equation}
  \label{d0pm} 
|d_0^+|^2 + |d_0^-|^2 \sim \frac{1}{Q}.  
\end{equation}

\section{Field enhancement}

In this section, we analyze the resonant effect of field enhancement
by a perturbation method. For a 2D periodic
structure given by a real dielectric function $\epsilon(x,y)$, we
assume there
is a resonant mode with a complex frequency $\omega_*$ and real Bloch
wavenumber $\beta$.  To avoid confusion with the diffraction solution
excited by incident waves, we denote the resonant mode by $u_*$, its
freespace wavenumber by $k_*$, define $\alpha_m^*$ by $\alpha_{m}^* = \sqrt{k_*^2 -
  \beta_m^2}$, but still denote the expansion coefficients of $u_*$ [as in
Eq.~(\ref{pwaves})] by $c_m^\pm$. The adjoint 
resonant mode is $v_*$, and its expansion coefficients are $d_m^\pm$. 

We consider a diffraction problem for incident waves with a 
real frequency $\omega$ near (or exactly at) the  real part of
$\omega_*$.  Two incident plane waves are given in the left ($x<-D$) and
right ($x>D$) of the periodic structure, respectively, and their
amplitudes $a_0^+$ and $a_0^-$ are assumed to satisfy
\begin{equation}
  \label{ampinc}
  |a_0^+|^2 + |a_0^-|^2 = 1.
\end{equation}
Since two incident waves are involved, we choose to normalize them 
to fix the total incident power. 
In the left and right homogeneous media, the total field can be
written as 
\begin{eqnarray}
  \label{dfexpand}
  &&  u({\bm r})  =  a_0^\pm e^{ i [\beta y \mp \alpha_0 (x \mp D)]}  \cr
  & & +  \sum_{m=-\infty}^\infty b_m^\pm  e^{ i [\beta_m y \pm 
        \alpha_m (x \mp D)]},
  \quad \pm x \ge  D, 
\end{eqnarray}
where $\alpha_m$ is defined in Eq.~(\ref{albeta}), 
and $b_m^\pm$ are the amplitudes of the outgoing plane waves. 
The wavevectors of the incident waves are $(\pm \alpha_0, \beta)$.
Notice that the resonant mode $u_*$ and the diffraction solution $u$ follow  
the same real Bloch wavenumber $\beta$.  
Again, we assume condition (\ref{1chanel}) is satisfied, then only $\alpha_0$ is
real positive and all $\alpha_m$ for 
$m\ne 0$ are pure imaginary with positive imaginary parts. 

To develop the perturbation theory, it is useful to write down the 
exact boundary conditions at $x=\pm D$ \cite{bao95}. 
Let $B$ be a linear operator acting on quasi-periodic functions
of $y$, such that 
\begin{equation}
  \label{defopB}
B e^{i \beta_m y} = i \alpha_m e^{ i \beta_m y}  
\end{equation}
for all integer $m$, then $u$ satisfies the following boundary
conditions
\begin{equation}
  \label{BCdf}
\pm  \frac{ \partial u}{\partial x}  = B u   -  2i \alpha_0 a_0^\pm e^{ i \beta y},
\quad x=\pm D.  
\end{equation}
For the complex frequency $\omega_*$, if we define a linear operator 
$B_*$ as in Eq.~(\ref{defopB}), with $\alpha_m$ replaced
by $\alpha^*_m$, then the resonant mode $u_*$ satisfies 
\begin{equation}
  \label{BCres}
\pm \frac{\partial u_*}{\partial x} = B_* u_*, \quad x= \pm D.
\end{equation}

If  $\delta = k - k_*$ is small (more precisely, $|\delta/k_*|$ is
small), we try to find the diffraction solution $u$ by the following
expansion: 
\begin{equation}
  \label{uexp1}
u = \frac{C}{\delta} u_* + u_0 + \delta u_1 + \delta^2 u_2  + \cdots  
\end{equation}
The operator $B$ must also be expanded: 
\begin{equation}
  \label{Bexp1}
B = B_* + \delta B_1 + \delta^2 B_2 + \cdots 
\end{equation}
It is easy to see that 
\[
\alpha_m = \sqrt{ k_*^2 - \beta_m^2 + k^2 - k_*^2} 
= \alpha^*_m  + \frac{k_* }{ \alpha^*_m} \delta + \cdots
\]
Therefore, $B_1$ is a linear operator satisfying 
\begin{equation}
  \label{defopB1}
B_1 e^{i \beta_m y} = 
\frac{ i k_*}{\alpha^*_m}  e^{
  i \beta_m y}   
\end{equation} 
for all integer $m$. 

Inserting the expansions for $u$ and $B$ into Eq.~(\ref{helm2d}) and
boundary condition (\ref{BCdf}), we collect equations and boundary
conditions at different powers of $\delta$. At $O(1/\delta)$, 
we simply get the Helmholtz equation and boundary conditions for $u_*$.
At $O(1)$, we obtain the following 
inhomogeneous Helmholtz equation
\begin{equation}
  \label{eq:u0}
\partial_x^2 u_0 + \partial_y^2 u_0 + k_*^2 \epsilon u_0 = -2 C k_*
\epsilon u_*
\end{equation}
and boundary conditions
\begin{equation}
  \label{bcD}
\pm \frac{\partial u_0}{\partial x}  - B_* u_0  =  C B_1 u_*  - 2i
\alpha_0^*  a_0^\pm 
 e^{i  \beta y},  \quad x=\pm D.
\end{equation}
Multiplying Eq.~(\ref{eq:u0}) by $v_*$ and integrating
on $\Omega$, we get 
\begin{equation}
  \label{Cform}
  C = \frac{iL \alpha_0^* \, (a_0^+ d_0^++ a_0^- d_0^- ) }
{k_* R}
\end{equation}
where
\[
R = \int_\Omega  \epsilon v_* u_* d{\bm r} + 
 \frac{iL}{2} \sum_{m=-\infty}^\infty 
\frac{ c_m^+ d_m^+ + c_m^- d_m^- }{ \alpha^*_m}.  
\]
Additional details are given in Appendix B. 

Field enhancement is  often
defined as the ratio of the amplitudes of the total and incident
waves. Since the incident waves are normalized according to
Eq.~(\ref{ampinc}), $u_*$ is scaled to satisfy Eq.~(\ref{max1}), and
$|\delta/k_*|$ is supposed to be small, the amplitude of $u$, and also
the field enhancement, can be
approximated by $|C/\delta|$. The term $a_0^+ d_0^+ + a_0^- d_0^-$
represents the coupling between the incident waves and the adjoint 
resonant mode $v_*$. If $(a_0^+, a_0^-)$ is proportional to 
$(d_0^-, -d_0^+)$, there is no field enhancement at all. 
To maximize $|C|$, we can choose the amplitudes of the 
incident  waves such that $(a_0^+, a_0^-)$ is proportional to 
$( \overline{d}_0^+, \overline{d}_0^-)$ 
then 
\[
| a_0^+ d_0^++ a_0^- d_0^- | = \sqrt{ |d_0^+|^2 + |d_0^-|^2}.
\]
The above is on the order of $1/\sqrt{Q}$. 
If $\omega = \mbox{Re}(\omega_*)$, then $\delta = k  - k_* = -
\mbox{Im}(k_*)$,  and thus $|C/\delta| \sim  \sqrt{Q}$.

\section{Numerical examples}

In this section, we present a number of numerical examples to
illustrate field enhancement near different kinds of BICs. The
periodic structure is an array of identical, parallel and infinitely
long circular cylinders. The cylinders are parallel to the $z$ axis,
arranged periodically along the $y$ axis with period $L$, and
surrounded by vacuum. The axis of one particular cylinder is exactly
the $z$ axis. The radius and dielectric constant of the cylinders are
$r_s$ and $\epsilon_s$, respectively. The structure has reflection
symmetry in both $x$ and $y$ directions.  For simplicity, we assume
there is only a single incident wave given in the left side of the
periodic structure, thus, $a_0^-=1$ and $a_0^+=0$.


The first example is an antisymmetric standing wave on a periodic array with
$\epsilon_s = 11.6$ and $r_s = 0.3L$. The frequency of this
symmetry-protected BIC is 
$\omega_\circ = 0.411227834 (2\pi c/L)$. 
Its electric field is
odd in $y$ (i.e., antisymmetric with respect to the reflection symmetry in
$y$) and even in $x$. First, we calculate a few resonant modes near this
BIC for some $\beta$ near $\beta_\circ = 0$. The complex frequencies  and
$Q$-factors of these resonant modes are listed in Table~\ref{T1a} below. 
\begin{table}[htp]
\caption{Example 1: Resonant modes near the BIC.}
\begin{center}
\begin{tabular}{c|c|c}
\hline 
 $\beta L/(2\pi )$ & $(\omega_{*}-\omega_\circ) L/(2\pi c)$  & $Q$-factor \\ \hline 
0.004 & -0.00001320  - 0.00000204i &   1.01$\times$$10^5$ \\
0.008 & -0.00005277  - 0.00000814i &   2.53$\times$$10^4$ \\
0.016 & -0.00021089  - 0.00003247i &   6.33$\times$$10^3$ \\
0.032 & -0.00084043  - 0.00012844i &   1.60$\times$$10^3$ \\
\hline 
\end{tabular}
\end{center}
\label{T1a}
\end{table}
It is easy to observe that
$\mbox{Re}(\omega_*) - \omega_\circ \sim \beta^2$, 
$\mbox{Im}(\omega_*) \sim \beta^2$ and 
$Q  \sim 1/\beta^2$. 

\begin{figure}[htb]
\centering
  \includegraphics[scale=0.8]{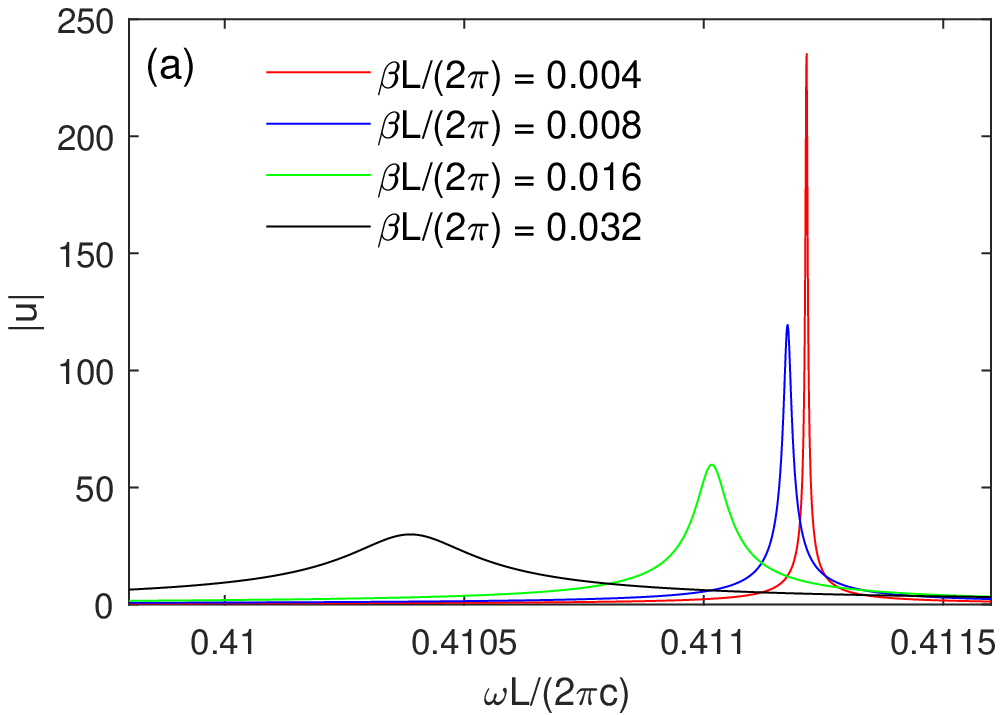}
  \includegraphics[scale=0.8]{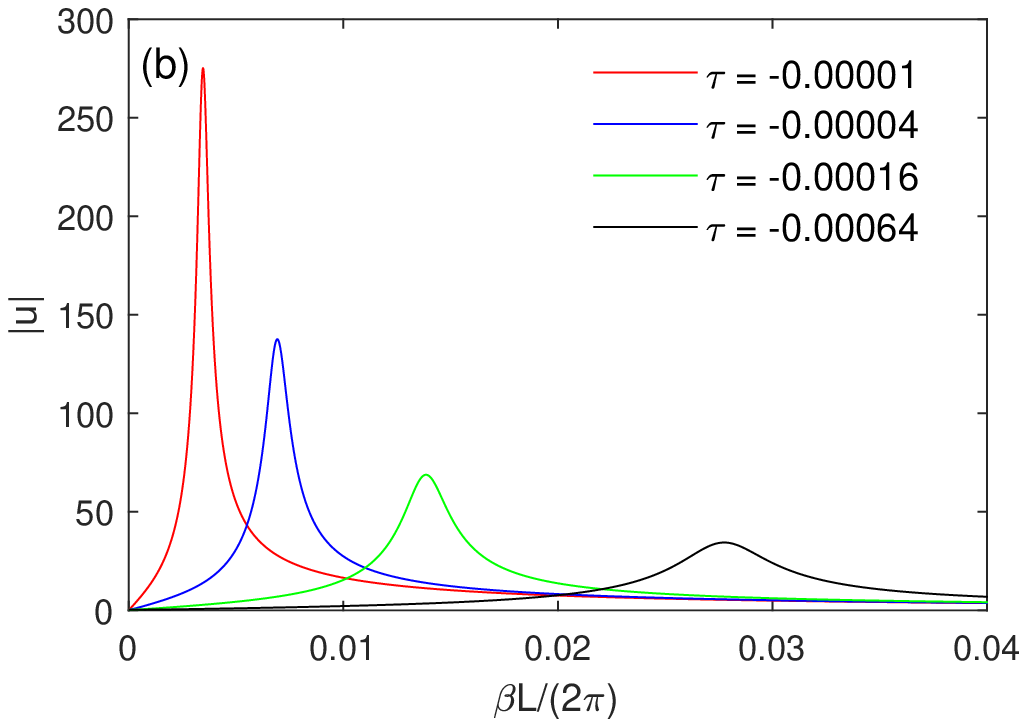}
\caption{Example 1: Magnitude of the electric field at point 
  $(0,0.2064L)$,  (a) as a function of $\omega$,  (b) as a function of 
  $\beta$. }
\label{E1f}
\end{figure}

Next, we solve the diffraction problem for incident plane waves 
with a real frequency and the same $\beta$ listed in
Table~\ref{T1a}. We monitor the solution at a particular point 
$(x,y)=(0, 0.2064L)$ for different frequencies. The results are
shown in Fig.~\ref{E1f}(a). 
For each $\beta$, we also find the maximum of $|u|$ over all
frequencies, and calculate the full width at half maximum (FWHM) $W_{\omega}$ for
$|u|$ as a function of $\omega$. 
 The results are listed in Table~\ref{T1b}. 
\begin{table}[htp]
\caption{Example 1: Maximum of $|u|$ as a function of $\omega$ for
  fixed $\beta$ and FWHM $W_{\omega}$.}
\begin{center}
\begin{tabular}{c|c|c}
\hline 
 $\beta L/(2\pi )$ & $\max_\omega |u|$ & $W_{\omega}$    \\
 \hline 
0.004  & 238.8 & 0.71$\times$$10^{-5}$ \\
0.008 & 119.4 & 2.82$\times$$10^{-5}$ \\
0.016 & 59.76 & 1.13$\times$$10^{-4}$ \\
0.032 & 29.96 & 4.45$\times$$10^{-4}$ \\
\hline 
\end{tabular}
\end{center}
\label{T1b}
\end{table}
The perturbation theory predicts that the 
maximum is reached when $\omega \approx \mbox{Re}(\omega_*)$. This is 
true to high accuracy when $Q$ is large. It is also easy to see that 
$\max_\omega |u| \sim 1/\beta$, and this confirms the perturbation result that 
enhancement should be proportional to  $1/\sqrt{Q}$.
The values of $W_\omega$ in Table~\ref{T1b} indicate that 
$W_{\omega}\sim\beta^2$. From the perturbation theory of Sec.~III,
Eq.~(\ref{uexp1}) in particular, we
know that the leading term of $u$ is inversely proportional to $\omega -
\mbox{Re}(\omega_*)$. Thus, the maximum is 
obtained when $\omega \approx \mbox{Re}(\omega_*)$, and 
half maximum is achieved when 
\begin{equation}
  \label{halfmax}
\omega \approx \mbox{Re}(\omega_*) \pm \sqrt{3} \, \mbox{Im}(\omega_*).  
\end{equation}
Therefore, $W_\omega \approx 2 \sqrt{3} \, |\mbox{Im}(\omega_*)| \sim
\beta^2$. 

We also study how the field depends on $\beta$ for a fixed
frequency near the BIC frequency $\omega_\circ$.  For this
BIC, the real part of 
$\omega_*$ is less than $\omega_\circ$ as shown in Table~\ref{T1a}.
Therefore,  we
consider the dependence on $\beta$ for $\omega$ slightly less than
$\omega_\circ$. The numerical results are shown in Fig.~\ref{E1f}(b),
where 
\begin{equation}
\tau = ( \omega - \omega_\circ ) L/(2\pi c).
\end{equation}
 We also calculate
$\beta_*$ and $W_\beta$, where $\beta_*$ is the value of $\beta$ that gives the
maximum of $|u|$, and $W_\beta$ is the FWHM for $|u|$ as a 
function of $\beta$. The results are listed
in Table~\ref{T1c}. 
\begin{table}[htp]
\caption{Example 1: Maximum of $|u|$ attained at $\beta_*$  for
  fixed $\omega$, and FWHM $W_\beta$.}
\begin{center}
\begin{tabular}{c|c|c|c}
\hline
 $(\omega - \omega_\circ)L/(2\pi c)$
 & $\beta_* L/(2\pi)$ & $\max_\beta |u| $ & $W_{\beta}$ \\
 \hline
-0.00001 & 0.00344390 & 275.2 & 0.92$\times$$10^{-3}$ \\
-0.00004 & 0.00690362  & 137.6 & 1.83$\times$$10^{-3}$ \\
-0.00016 & 0.01381251  & 68.80 & 3.67$\times$$10^{-3}$ \\
-0.00064 & 0.02766718 & 34.24 &  7.34$\times$$10^{-3}$  \\ \hline
\end{tabular}
\end{center}
\label{T1c}
\end{table}
Since the leading term of $u$ is proportional to $1/[\omega -
\omega_*(\beta)]$, $\beta_*$ should satisfy 
$\omega = \mbox{Re}[ \omega_*(\beta_*)]$ approximately. As
$\mbox{Re}[ \omega_*(\beta)] - \omega_\circ$ depends on $\beta$
quadratically, we easily obtain $\beta_* \sim  |\omega - \omega_\circ|^{1/2}$. The
maximum of $|u|$ 
is proportional to $\sqrt{Q}$ for that $\beta_*$, and thus it is
proportional to $ | \omega - \omega_\circ|^{-1/2}$. The two $\beta$
values at half maximum
can be approximately calculated from the following equation
\begin{equation}
  \label{halfmax2}
| \omega - \omega_*(\beta)| = 2 | \mbox{Im}[\omega_*(\beta_*)] |  
\end{equation}
Using the leading order approximation for $\omega_*(\beta) -
\omega_\circ$, it is easy to show that both solutions of 
Eq.~(\ref{halfmax2}), as well as their difference, scale as
$| \omega - \omega_\circ |^{1/2}$. Therefore, $W_{\beta} \sim | \omega
- \omega_\circ |^{1/2}$. 
All these asymptotic relations  are confirmed by the
numerical results listed in Table~\ref{T1c}. 


The second example is a symmetric standing wave on a periodic array of
cylinders with dielectric constant $\epsilon_s =10$ and radius
$r_s = 0.36665158L$. The frequency of this BIC is
 $ \omega_\circ  =  0.491142367 (2\pi c/L)$. 
It is not
a symmetry-protected BIC, since its field pattern is symmetric in $y$
(i.e. $u$ is even in $y$). It turns out that the BIC is also even
in $x$. In Table~\ref{T2a}, 
\begin{table}[htp]
\caption{Example 2: Resonant modes near the BIC.}
\begin{center}
\begin{tabular}{c|c|c}
\hline 
 $\beta L/(2\pi)$ & $(\omega_{*}-\omega_\circ) L/(2\pi c)$ & $Q$ \\
 \hline 
0.01 &                  0.000093965 - 0.000000025i&  9.77$\times$$10^6$\\
$0.01\sqrt{2}$ &  0.000187712 - 0.000000093i&  2.63$\times$$10^6$\\
0.02 &                  0.000374556 - 0.000000358i & 6.87$\times$$10^5$\\
$0.02\sqrt{2}$ &  0.000745686 - 0.000001393i & 1.77$\times$$10^5$\\
\hline 
\end{tabular}
\end{center}
\label{T2a}
\end{table}
we list the complex frequencies and $Q$-factors for a few resonant
modes near the BIC. These results confirm that 
$\mbox{Re}(\omega_*) -\omega_\circ \sim \beta^2$, 
$\mbox{Im}(\omega_*)  \sim \beta^4$, and 
$Q \sim 1/\beta^4$. 

\begin{figure}[htb]
  \centering
  \includegraphics[scale=0.8]{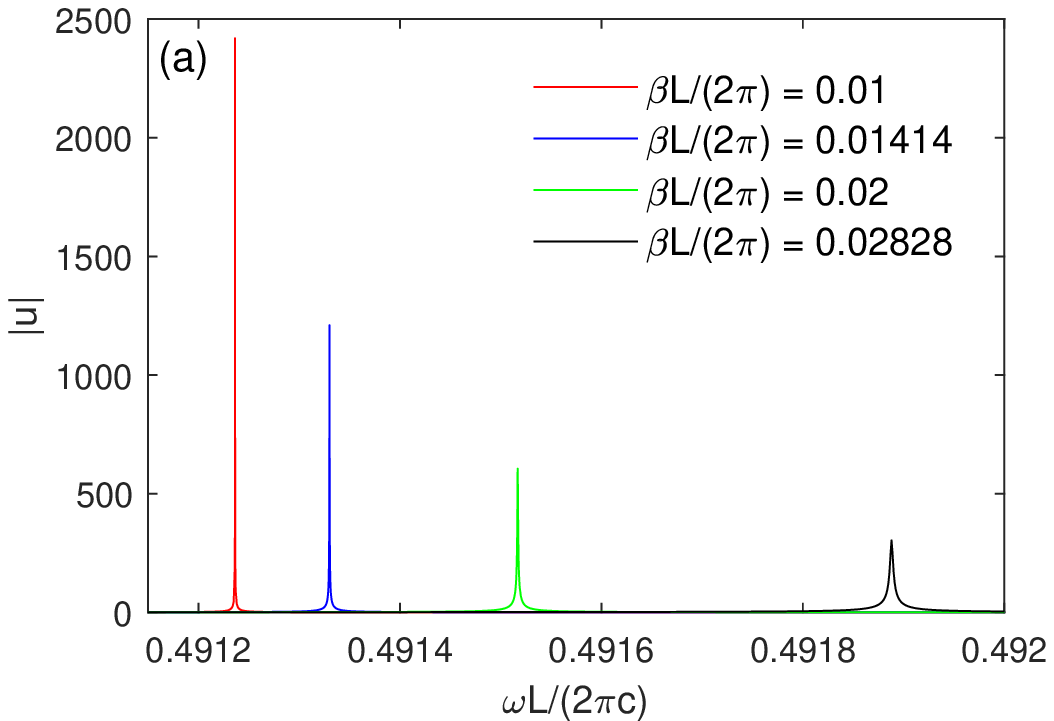}
  \includegraphics[scale=0.8]{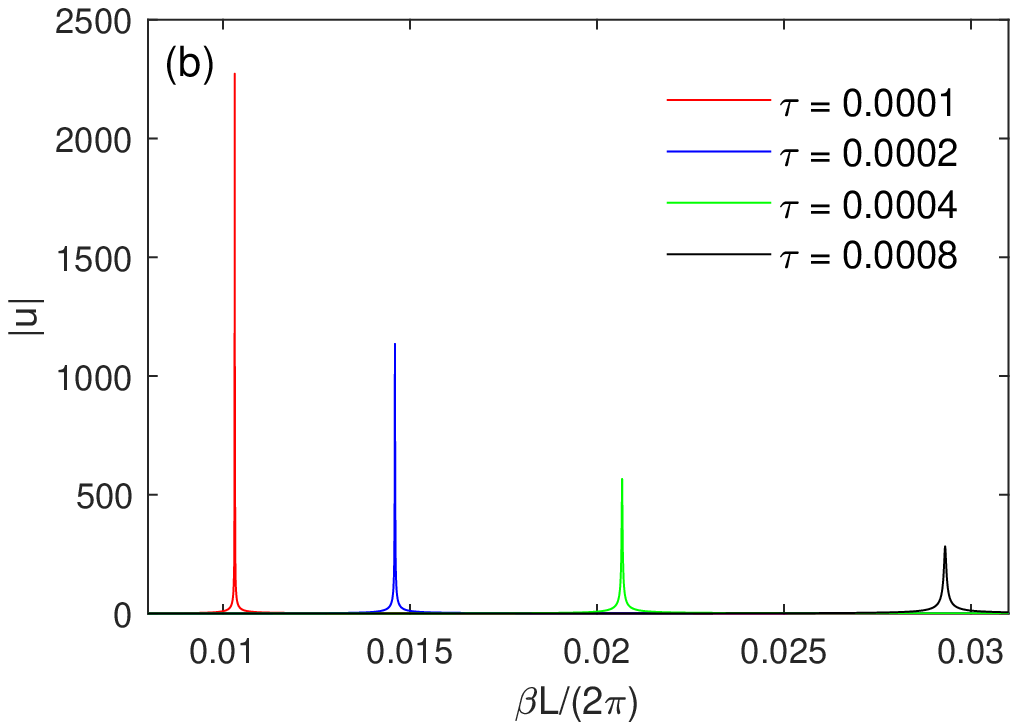}
\caption{Example 2: Magnitude of the electric field at 
point $(0, -0.3462L)$,  (a) as a function of $\omega$, (b) as 
a function of $\beta$.}
\label{E2f}
\end{figure}

Next, we solve the diffraction problem with a plane incident wave, and
monitor the solution at a particular point $(x,y)=(0, -0.3462L)$. 
In Fig.~\ref{E2f}(a) and (b), 
we show $|u|$ at that point as a function of $\omega$ for fixed $\beta$
and as a function of $\beta$ for fixed $\omega$, respectively. For the
case of fixed $\beta$, the maximum of $|u|$ and FWHM $W_\omega$ are
listed in Table~\ref{T2b}. 
\begin{table}[htp]
\caption{Example 2: Maximum of $|u|$ as a function of $\omega$ for fixed 
  $\beta$ and FWHM $W_\omega$.}
\begin{center}
\begin{tabular}{c|c|c}
\hline 
 $\beta L/(2\pi)$ & $\max_\omega |u|$ & $W_{\omega}$   \\
 \hline 
0.01 & 2419 &  0.87$\times$$10^{-7}$ \\
$0.01\sqrt{2}$ & 1210 & 3.23$\times$$10^{-7}$ \\
0.02 &  605.5 &   1.24$\times$$10^{-6}$  \\
$0.02\sqrt{2}$ & 303.2 & 4.82$\times$$10^{-6}$ \\
\hline 
\end{tabular}
\end{center}
\label{T2b}
\end{table}
These numerical results indicate that $\max |u| \sim 1/\beta^2$ 
and $W_\omega \sim \beta^4$, and they support our claims that 
field enhancement is proportional to $\sqrt{Q}$ and 
$W_\omega \approx 2\sqrt{3} \, | \mbox{Im}(\omega_*) |$. From
Table~\ref{T2a}, we see that $\mbox{Re}(\omega_*)$ is larger than
$\omega_\circ$ for this BIC. Therefore, we show $|u|$ as functions of
$\beta$ for $\omega$ slightly larger than $\omega_\circ$ in
Fig.~\ref{E2f}(b). For each fixed $\omega$, we list $\beta_*$,
$\max_\beta |u|$ and FWHM $W_\beta$ in Table~\ref{T2c}. 
\begin{table}[htp]
\caption{Example 2: Maximum of $|u|$ attained at $\beta_*$ for fixed 
  $\omega$, and FWHM $W_\beta$.}
\begin{center}
\begin{tabular}{c|c|c|c}
\hline
 $(\omega -\omega_\circ)L/(2\pi c)$ & $\beta_{*} L/(2\pi )$ &
                                                              $\max_\beta |u|$ & $W_{\beta}$    \\
 \hline
0.0001 & 0.01031652 & 2273 & 0.51$\times$$10^{-5}$ \\
0.0002 & 0.01459880 & 1136 & 1.34$\times$$10^{-5}$ \\
0.0004 & 0.02067137 & 566.9 & 3.66$\times$$10^{-5}$ \\
0.0008 & 0.02930596 & 282.5 & 1.03$\times$$10^{-4}$ \\
\hline
\end{tabular}
\end{center}
\label{T2c}
\end{table}
Since the maximum of $|u|$ is approximately attained when $\omega =  
\mbox{Re}(\omega_*)$, we have $\beta_* \sim  |\omega -
  \omega_\circ|^{1/2}$. This also implies that $\max_\beta |u| \sim \sqrt{Q}
\sim 1/|\omega - \omega_\circ|$. 
To estimate $W_\beta$, we assume 
the equation $|\omega - \omega_*(\beta)| = 2 | \mbox{Im}[
\omega_*(\beta_*)]|$ has two solutions near $\beta_*$. The Taylor
expansion at $\beta_*$ gives 
\[
\omega - \omega_*(\beta) = - \mbox{Im}[ \omega_*(\beta_*)] -
\omega_*'(\beta_*) 
(\beta-\beta_*) + ...
\]
where $\omega_*'$ is the derivative of $\omega_*$ with respect to
$\beta$. Therefore, the two terms in the right hand side above must have the
same order. Since $\mbox{Im}[ \omega_*(\beta_*)] = O(\beta_*^3)$ and 
$\omega_*'(\beta_*)  = O(\beta_*)$, we conclude that 
$\beta- \beta_* = O(\beta_*^3)$. This leads to $W_\beta =
O(\beta_*^3)$ or $W_\beta  \sim |\omega - \omega_\circ|^{3/2}$. The
numerical results of Table~\ref{T2c} are consistent with these asymptotic
relations. In particular, the last column of Table~\ref{T2c} confirm
that when $|\omega - \omega_\circ|$ is increased 
by a factor of $2$, $W_\beta$ is increased by a factor of $2^{1.5}
\approx 2.83$. 


The third example is a propagating BIC on a periodic array of
cylinders with $\epsilon_s = 11.56$ and $r_s = 0.35L$. The frequency
and Bloch wavenumber of the BIC are 
$\omega_\circ = 0.670236140 (2\pi c/L)$
 and 
$\beta_\circ = 0.2483 (2\pi/L)$, respectively. In Table~\ref{T3a},
\begin{table}[htp]
\caption{Example 3: Resonant modes near the BIC.}
\begin{center}
\begin{tabular}{c|c|c}
\hline 
 $(\beta-\beta_\circ ) L/(2\pi )$ & 
$(\omega_{*}-\omega_\circ) L/(2\pi c)$  & $Q$-factor \\ 
 \hline 
0.004 & 0.00025157 - 0.00000127i  &  2.63$\times$$10^5$ \\
0.008 & 0.00049981 - 0.00000518i &  6.47$\times$$10^4$\\
0.016 & 0.00098439 - 0.00002120i &  1.58$\times$$10^4$\\
0.032 & 0.00188950 - 0.00008790i &  3.82$\times$$10^3$\\
\hline 
\end{tabular}
\end{center}
\label{T3a}
\end{table}
we show the complex frequencies and $Q$-factors of a few resonant modes
for $\beta$ near $\beta_\circ$. For this BIC, it is clear that 
$\mbox{Re}(\omega_*) - \omega_\circ  \sim \beta - \beta_\circ$, 
$\mbox{Im}(\omega_*) \sim (\beta - \beta_\circ)^2$, and 
$Q \sim 1/(\beta - \beta_\circ)^2$. 

\begin{figure}[htb]
  \centering
  \includegraphics[scale=0.8]{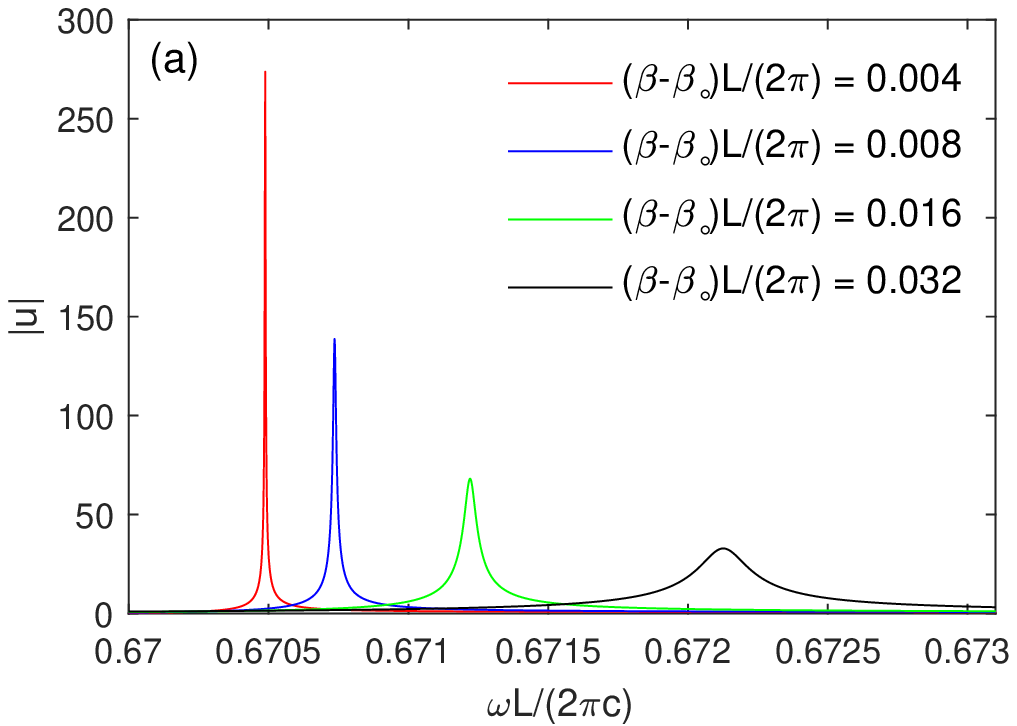}
  \includegraphics[scale=0.8]{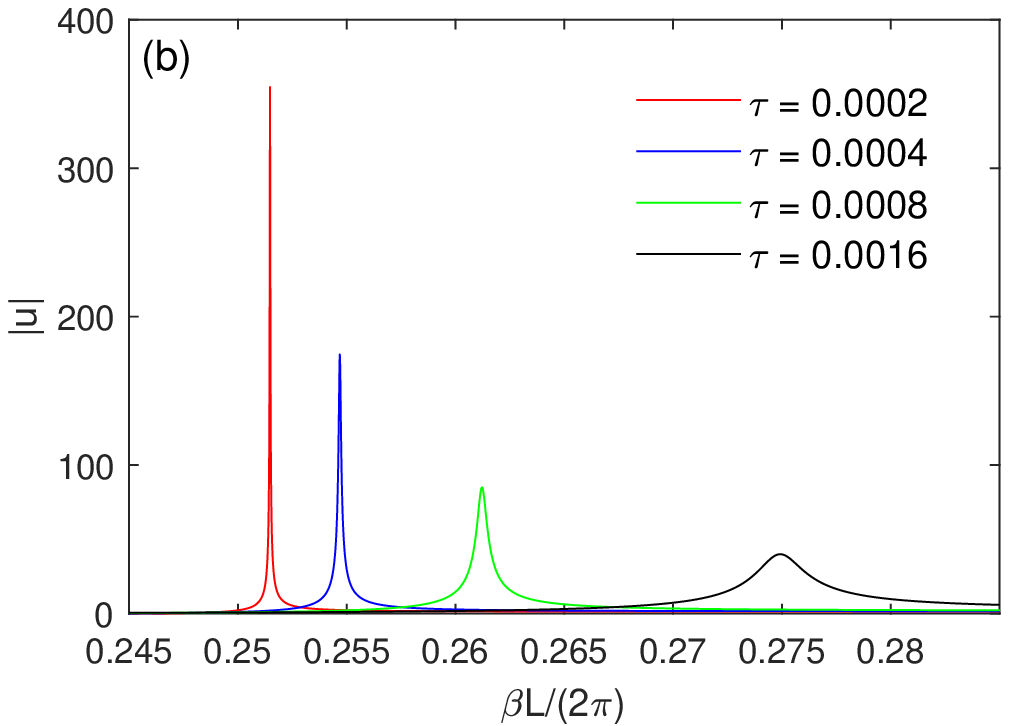}
\caption{Example 3: Magnitude of the electric field at point 
  $(0.1526L, -0.2579L)$, (a) as a function of $\omega$, (b) as a function of 
  $\beta$.}
\label{E3f}
\end{figure}

For the diffraction problem with an incident wave of unit amplitude,
we monitor the solution at point $(x,y)=(0.1526L, -0.2579L)$.
In Fig.~\ref{E3f}(a) and (b), 
we show $|u|$ at that point as functions of $\omega$ or $\beta$,
respectively. For a few fixed values of $\beta-\beta_\circ$, the
maximum of $|u|$ and FWHM $W_\omega$ are
listed in Table~\ref{T3b}.
\begin{table}[htp]
\caption{Example 3: Maximum of $|u|$ as a function of $\omega$ for fixed 
  $\beta$ and FWHM $W_\omega$.}
\begin{center}
\begin{tabular}{c|c|c}
\hline 
 $(\beta-\beta_\circ) L/(2\pi )$ & $\max_\omega |u|$ & $W_\omega $ \\
 \hline 
0.004 &   280.6 & 4.41$\times$$10^{-6}$  \\
0.008 &  138.7 & 1.80$\times$$10^{-5}$  \\
0.016 &  68.10 & 7.40$\times$$10^{-5}$  \\
0.032 &  32.87 & 3.05$\times$$10^{-4}$  \\
\hline 
\end{tabular}
\end{center}
\label{T3b}
\end{table}
These numerical results indicate that $\max_\omega |u|
\sim 1 /|\beta-\beta_\circ|$ and 
$W_\omega \sim |\beta - \beta_\circ|^2$, and they are consistent with
the results on field enhancement and 
$W_\omega \approx 2 \sqrt{3}\, \mbox{Im}(\omega_*) \sim
(\beta-\beta_\circ )^2$. For a few fixed frequencies near
$\omega_\circ$, we list $\max_\beta |u|$, $\beta_*$ and $W_\beta$ in
Table~\ref{T3c}. 
\begin{table}[h!]
\caption{Example 3: Maximum of $|u|$ attained at $\beta_*$ for fixed 
  $\omega$, and FWHM $W_\beta$.}
\begin{center}
\begin{tabular}{c|c|c|c}
\hline 
 $(\omega -\omega_\circ )L/(2\pi c)$ & $\beta_* L/(2\pi )$ &
                                                             $\max_\beta |u|$ & $W_\beta$    \\
 \hline  
0.0002 & 0.25147581  & 354.6 & 4.41$\times$$10^{-5}$ \\ 
0.0004 & 0.25468389  & 174.5 & 1.84$\times$$10^{-4}$ \\ 
0.0008 & 0.26121366 & 84.93 & 7.87$\times$$10^{-4}$ \\ 
0.0016 & 0.27488676 & 39.93 & 3.75$\times$$10^{-3}$ \\ 
\hline 
\end{tabular}
\end{center}
\label{T3c}
\end{table} 
The maximum of $|u|$ is attained at $\beta_*$ which satisfies 
$\omega = \mbox{Re}[ \omega_*(\beta_*)]$ approximately. 
Therefore, $\beta_* - \beta_\circ \sim \omega - \omega_\circ$. In
addition, $\max_\beta |u|$ should be proportional to $\sqrt{Q}$ for
the corresponding $\beta_*$. Therefore, $\max_\beta|u| \sim 1/|\beta_*
- \beta_\circ| \sim 1/|\omega - \omega_\circ|$. 
To determine $W_\omega$, we first estimate the $\beta$ that gives
half maximum. As before, we know that $\mbox{Im}[ \omega_*(\beta_*)]$ and 
$\omega_*'(\beta_*)  (\beta-\beta_*)$ should be on the same order, but
$\omega_*'(\beta_*)]$ is a nonzero constant, thus 
$  \beta-\beta_\circ \sim (\beta_* - \beta_\circ)^2 \sim (\omega -
\omega_\circ)^2$. Therefore, 
$W_{\beta}\sim  (\omega -\omega_\circ )^2$.


The fourth example is an antisymmetric standing wave on a periodic
array with $\epsilon_s = 8.2$  and $r_s = 0.432266L$.  The frequency
of this BIC is 
$\omega_\circ = 0.770094460005 (2\pi c/L)$. 
In Table~\ref{T4a}, 
\begin{table}[htp]
\caption{Example 4: Resonant modes near the BIC.}
\begin{center}
\begin{tabular}{c|c|c}
\hline 
 $\beta L/(2\pi )$ & $(\omega_*-\omega_\circ) L/(2\pi c)$ & $Q$-factor \\
 \hline 
0.005 & -0.000036385554 - 0.000000000024i &  1.63$\times$$10^{10}$\\
0.01 &   -0.000145089692 - 0.000000001751i  &   2.20$\times$$10^{8}$\\
0.02 &   -0.000573518292 - 0.000000102698i &  3.75$\times$$10^{6}$\\
0.04 &   -0.002204078427 - 0.000004284774i &   8.96$\times$$10^{4}$\\
\hline  
\end{tabular}
\end{center}
\label{T4a}
\end{table}
we list the complex frequencies and $Q$-factors of a few resonant modes
with $\beta$ close to $\beta_\circ = 0$. It is quite clear that 
$\mbox{Re}(\omega_*) -\omega_\circ \sim \beta^2$, 
 $\mbox{Im}(\omega_*) \sim \beta^6$, and $Q \sim 1/\beta^6$.

\begin{figure}[htb]
  \centering 
  \includegraphics[scale=0.75]{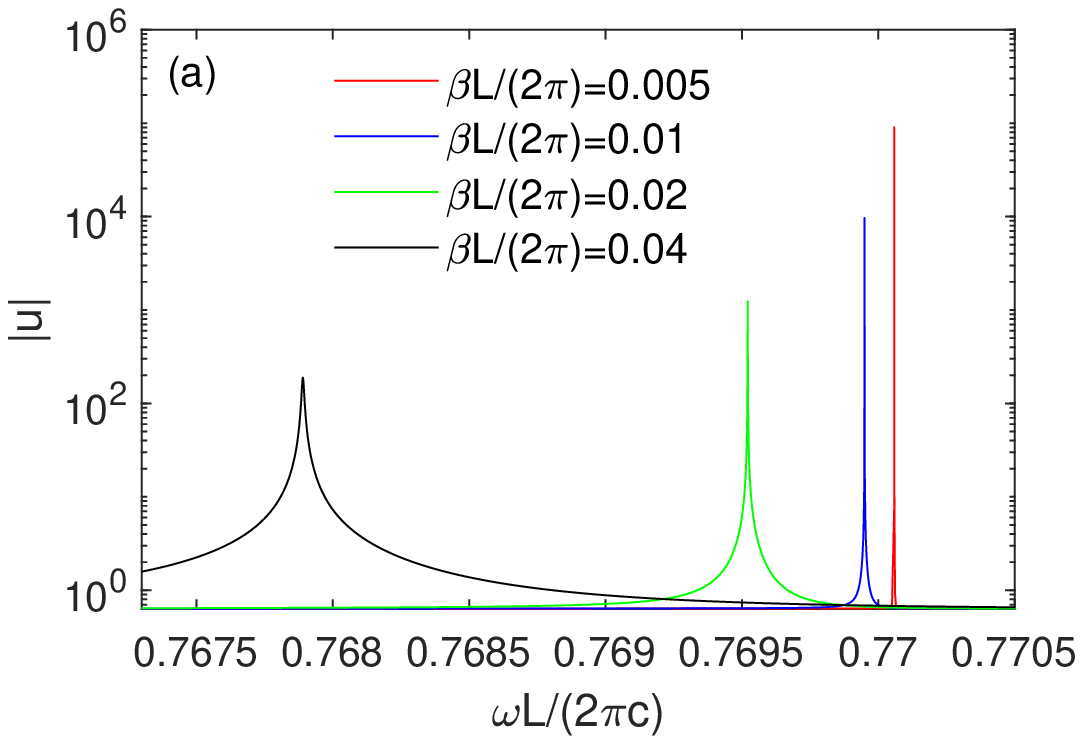}
  \includegraphics[scale=0.75]{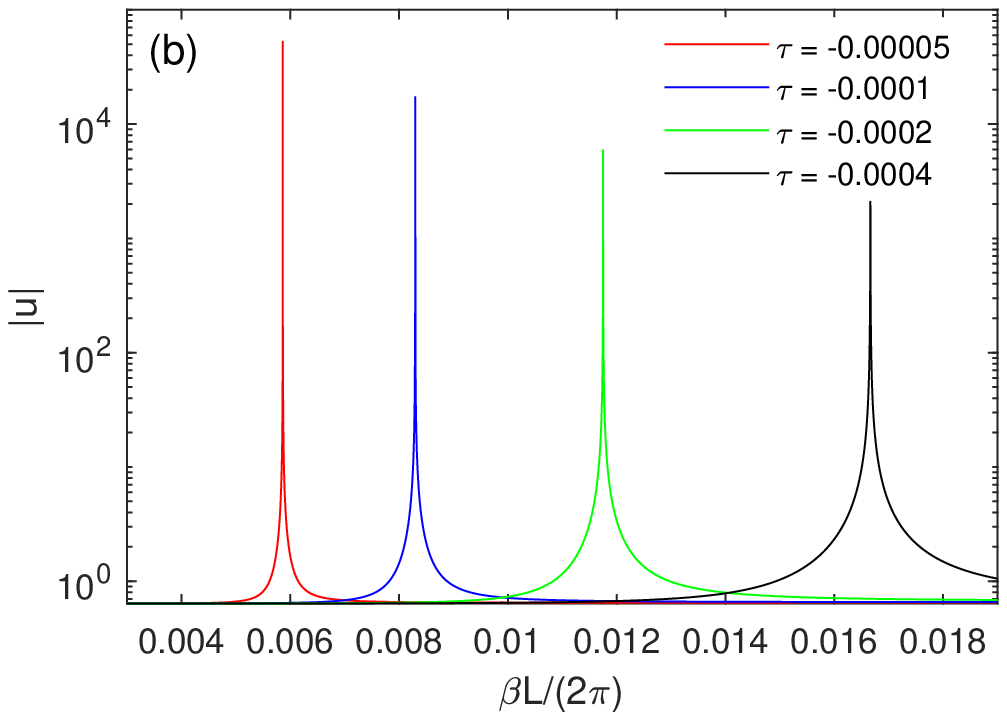}
\caption{Example 4: Magnitude of the electric field at 
point $(0.1237L, 0)$,  (a) as a function of $\omega$, (b) as 
a function of $\beta$.}
\label{E4f}
\end{figure}

For the diffraction problem, we monitor the solution at $(x,y)= (0.1237L, 0)$. 
The numerical results are shown in Fig.~\ref{E4f} 
for fixed $\beta$ near $\beta_\circ =0$ and fixed $\omega$ near $\omega_\circ$. 
The maximum of $|u|$ for fixed $\beta$ are listed in Table~\ref{T4b}. 
\begin{table}[htp]
\caption{Example 4: Maximum of $|u|$ as a function of $\omega$ for fixed 
  $\beta$ and FWHM $W_\omega$.}
\begin{center}
\begin{tabular}{c|c|c}
\hline 
 $\beta L/(2\pi )$ & $\max_\omega |u|$ & $W_\omega$ \\
 \hline 
0.005 &  9.03$\times$$10^{4}$ & 8.2$\times$$10^{-11}$ \\
0.01 &   9.67$\times$$10^3$ & 6.1$\times$$10^{-9}$ \\
0.02 &  1.24$\times$$10^3$ & 3.6$\times$$10^{-7}$ \\
0.04 &  1.89$\times$$10^2$ & 1.5$\times$$10^{-5}$ \\
\hline 
\end{tabular}
\end{center}
\label{T4b}
\end{table}
Since $\mbox{Im}(\beta_*) \sim \beta^6$, we have $Q \sim 1/\beta^6$, 
$\max_\omega |u| \sim 1/\sqrt{Q} \sim 1/\beta^3$, 
and $W_\omega \approx 2\sqrt{3} \mbox{Im}(\omega_*) \sim \beta^6$.  In
Table~\ref{T4c},  
\begin{table}[htp]
\caption{Example 4: Maximum of $|u|$ attained at $\beta_*$ for fixed 
  $\omega$, and FWHM $W_\beta$.}
\begin{center}
\begin{tabular}{c|c|c|c}
\hline 
 $(\omega  -\omega_\circ )L/(2\pi c)$ & $\beta_* L/(2\pi )$ &
                                                              $\max_\beta |u| $ & $W_{\beta}$    \\
 \hline 
-0.00005&0.005862404 & 5.26$\times$$10^4$ & 1.3$\times$$10^{-8}$ \\
-0.0001 & 0.008296643 & 1.72$\times$$10^4$ & 8.1$\times$$10^{-8}$ \\
-0.0002 & 0.011749913 & 5.92$\times$$10^3$ & 4.7$\times$$10^{-7}$ \\
-0.0004 & 0.016663279 & 2.10$\times$$10^3$ & 2.6$\times$$10^{-6}$ \\
\hline 
\end{tabular}
\end{center}
\label{T4c}
\end{table}
we list the maximum of $|u|$ for fixed $\omega$ slightly smaller than
$\omega_\circ$.  From the result on the real part of $\omega_*$, it is
easy to show that $\beta_* \sim |\omega -
\omega_\circ|^{1/2}$. Meanwhile,  $\max_\beta |u|$ should be proportional to 
$\beta_*^{-3}$ or $|\omega - \omega_\circ|^{-1.5}$. As before, the two
$\beta$ that reach half the maximum satisfy $\omega_*'(\beta_*) (
\beta - \beta_*)  \sim \mbox{Im}[\omega_*(\beta_*)] \sim \beta_*^6$. 
Since $\omega_*'(\beta_*) \sim \beta_*$, then both $|\beta - \beta_*|$
and $W_\beta$ are $O( \beta_*^5)$. Therefore, $W_\beta \sim |\omega -
\omega_\circ|^{2.5}$. The numerical results in Tables~\ref{T4b} and
\ref{T4c} confirm all these asymptotic results. 

\section{Conclusions}

Field enhancement by high-$Q$ resonances is crucial for realizing many
applications in photonics. Since a BIC on a periodic structure is 
surrounded  by resonant modes with $Q$-factors approaching infinity,
it is important to develop asymptotic formulas for field enhancement
around BICs. In this paper, we derived a formula for resonant field
enhancement  on 2D periodic structures (with 1D periodicity)
sandwiched between two homogeneous media,  and performed accurate numerical
calculations for field enhancement around some BICs exhibiting different
asymptotic relations. Although our study is
for 2D structures, we expect the results still hold for 3D biperiodic
structures such as photonic crystal slabs. Instead of varying the
Bloch wavenumber, high-$Q$ resonant modes can also be created by
perturbing the structure. The theory on field
enhancement developed in Sec.~III is also applicable to these
resonances. 

In practice, a small material loss is always present in any dielectric
material, and it sets a limit for both the $Q$-factor and the field
enhancement \cite{yoon15}. The material loss also has nontrivial effects on
some BICs without symmetry protection \cite{huyuan19}. Further studies are
needed to estimate the $Q$-factors and field enhancement for
realistic structures that are finite, nonperiodic and lossy, and with 
fabrication errors that destroy the relevant symmetries.

\section*{Acknowledgments}
The authors acknowledge support from 
the Fundamental Research Funds for the Central 
Universities of China (Grant No. 2018B19514), 
 the Science and Technology 
Research Program of Chongqing Municipal Education Commission, China 
(Grant No. KJ1706155), and 
the Research Grants Council of 
Hong Kong Special Administrative Region, China (Grant No. CityU 
11304117). 

\section*{Appendix A}
Let $u$ be a resonant mode satisfying Eqs.~(\ref{helm2d}) and
(\ref{pwaves}). Multiplying $\overline{u}$ to both sides of
Eq.~(\ref{helm2d}), integrating on $\Omega$ and using integration by parts, we have 
\begin{equation}
  \label{app1}
\int_{\partial \Omega} \overline{u} \frac{\partial u}{\partial \nu} ds 
- \int_\Omega |\nabla u|^2 d{\bm r} + k^2 
\int_\Omega \epsilon |u|^2 d{\bm r} = 0,
\end{equation}
where $\partial \Omega$ is the boundary of $\Omega$, $\nu$ is the
outward unit normal vector of $\Omega$. Due to the quasi-periodic
condition in $y$, the line integrals at $y=\pm L/2$ cancel, thus 
\[
\int_{\partial \Omega} \overline{u} \frac{\partial u}{\partial \nu} ds 
= \int_{-L/2}^{L/2} 
 \left[  \overline{u} \frac{\partial u}{\partial x}
  \right]^{x=D}_{x=-D}   dy,
\]
where $[F(x,y)] ^{x=D}_{x=-D} = F(D,y) - F(-D,y)$. 
Evaluating the right hand side above using Eq.~(\ref{pwaves}), we
obtain 
\begin{equation}
  \label{app2}
\int_{\partial \Omega} \overline{u} \frac{\partial u}{\partial \nu} ds 
= L \sum_{m=-\infty}^\infty i \alpha_m (|c_m^+|^2 + |c_m^-|^2).  
\end{equation}
Taking the imaginary parts of Eqs.~(\ref{app1}) and (\ref{app2}), we
have 
\[
L \sum_{m}   (|c_m^+|^2 + |c_m^-|^2)
\mbox{Re}(\alpha_m)  
+  \mbox{Im}(k^2) 
\int_\Omega \epsilon |u|^2 d{\bm r} = 0.
\]

Let $\Omega_e^+$ be the domain given by $x>D$ and $|y|<L/2$, and $u_e$ be
the sum of all terms with $m\ne 0$ in Eq.~(\ref{pwaves}), then
$u_e$ satisfies 
\[
( \partial_x^2 + \partial_y^2 + k^2 )  u_e = 0.
\]
Multiplying the above by $\overline{u}_e$ and integrating on
$\Omega_e^+$, we get 
\[
L \sum_{m\ne 0} |c_m^+|^2 \mbox{Re}(\alpha_m) = \mbox{Im}(k^2) 
\int_{\Omega_e^+} |u_e|^2 d{\bm r}.
\]
A similar result holds for $\Omega_e^-$ given by $x < -D$ and
$|y|<L/2$, then for $\Omega_e = \Omega_e^+ \cup \Omega_e^-$, we have 
\[
L \sum_{m\ne 0} ( |c_m^+|^2 + |c_m^-|^2)  \mbox{Re}(\alpha_m) = \mbox{Im}(k^2) 
\int_{\Omega_e} |u_e|^2 d{\bm r}.
\]
Combining the above equations, we obtain 
\begin{eqnarray*}
&& - \mbox{Im}(k^2) \left(
\int_{\Omega} \epsilon |u|^2 d{\bm r}
+ \int_{\Omega_e^+} |u_e|^2 d{\bm r} 
\right) \cr 
&& =  L (|c_0^+|^2 + |c_0^-|^2) \mbox{Re}(\alpha_0). 
\end{eqnarray*}
Noticing that $\mbox{Im}(k^2) = 2 \mbox{Re}(k) \mbox{Im}(k)$ 
and $1/Q = -2 \mbox{Im}(k)/\mbox{Re}(k)$, the above leads to
Eq.~(\ref{ourQ}).

\section*{Appendix B}
Multiplying Eq.~(\ref{eq:u0}) by $v_*$, integrating on $\Omega$,
we get
\begin{equation}
  \label{app3}
\int_{\partial \Omega} 
\left(  v_* \frac{\partial u_0}{\partial \nu} -
 u_0  \frac{\partial v_*} {\partial \nu} \right) ds 
= -2 C k_* \int_\Omega \epsilon u_* v_* d{\bm r}.  
\end{equation}
In the above, we used Green's second identity and noticed that 
$v_*$ satisfies the same Helmholtz equation as $u_*$. For the left
hand side above, the line integrals at $y=\pm L/2$ cancel out, thus 
\begin{eqnarray*}
&&
\int_{\partial \Omega} 
  v_* \frac{\partial u_0}{\partial \nu} ds 
= \int_{-L/2}^{L/2} 
\left[  v_* \frac{\partial u_0}{\partial x} \right]^{x=D}_{x=-D}
dy, \cr
&& \int_{\partial \Omega} 
 u_0  \frac{\partial v_*} {\partial \nu}  ds 
 = \int_{-L/2}^{L/2} 
\left[  u_0  \frac{\partial v_*} {\partial x} \right]^{x=D}_{x=-D}
dy.
\end{eqnarray*}
For $\partial_x u_0$, we use the boundary condition (\ref{bcD}). For
$v_*$ and $\partial_x v_*$, we use the expansion (\ref{vpwave}).  It is
easy to verify that 
\[
\int_{-L/2}^{L/2} \left[  v_* (B_* u_0) - u_0  \frac{\partial v_*}
  {\partial x} \right]_{x=D} dy = 0.
\]
Thus, 
\begin{eqnarray*}
&& \int_{-L/2}^{L/2} \left[  v_* \frac{\partial u_0}{\partial x} 
- u_0  \frac{\partial v_*} {\partial x} \right]_{x=D} dy \cr
&& =  \int_{-L/2}^{L/2} \left[  C  v_* (B_1 u_*) 
 - 2 i \alpha_0^* a_0^+      e^{i \beta    y}
   v_* \right]_{x=D}   dy \cr
&& = 
i L k_* C \sum_{m} c_m^+ d_m^+ /\alpha_m^* - 
2 i L  \alpha_0^* a_0^+ d_0^+.
\end{eqnarray*}
A similar result holds for the line integral at $x=-D$. Therefore, 
\begin{eqnarray}
&& \int_{\partial \Omega} 
\left(  v_* \frac{\partial u_0}{\partial \nu} -
 u_0  \frac{\partial v_*} {\partial \nu} \right) ds  
 + 2 i L  \alpha_0^*  (a_0^+ d_0^+  + a_0^- d_0^-) \cr
&& 
\label{app4}
 =  i L k_* C \sum_{m=-\infty}^\infty  \frac{ c_m^+ d_m^+ + c_m^- d_m^-}{\alpha_m^*}.
\end{eqnarray}
Inserting Eq.~(\ref{app3}) to Eq.~(\ref{app4}), we obtain Eq.~(\ref{Cform}).


\begin{thebibliography}{99}

\bibitem{hsu16} C. W. Hsu, B. Zhen, A. D. Stone, 
  J. D. Joannopoulos, and M. Solja\v{c}i\'{c}, 
 ``Bound states in the continuum,'' 
Nat. Rev. Mater. {\bf 1}, 16048 (2016). 

\bibitem{kosh19} K. Koshelev, G. Favraud, A. Bogdanov, Y. Kivshar, 
and A. Fratalocchi, ``Nonradiating photonics with resonant dielectric 
nanostructures,'' Nanophotonics {\bf 8}, 725--745 (2019). 

\bibitem{neumann29}  J. von Neumann and E. Wigner, 
 ``\"{U}ber   merkw\"{u}rdige diskrete eigenwerte,'' 
Z. Physik  {\bf 50},  291-293 (1929). 

 \bibitem{bonnet94} A.-S. Bonnet-Bendhia and F. Starling, ``Guided 
 waves by electromagnetic gratings and nonuniqueness examples for the 
 diffraction problem,''  Math. Methods Appl. Sci.  {\bf 17}, 305-338 
 (1994).  

\bibitem{padd00} P. Paddon and J. F. Young, ``Two-dimensional 
  vector-coupled-mode theory for textured planar waveguides,'' 
\prb\ {\bf 61}, 2090-2101 (2000). 

\bibitem{tikh02} S. G. Tikhodeev, A. L. Yablonskii, E. A Muljarov, 
  N. A. Gippius, and T. Ishihara, 
``Quasi-guided modes and optical properties of photonic crystal 
slabs,'' \prb\ {\bf 66}, 045102 (2002). 


\bibitem{shipman03} S. P. Shipman and S. Venakides, 
``Resonance and bound states in photonic crystal slabs,'' 
SIAM J. Appl. Math.  {\bf 64}, 322-342 (2003). 

\bibitem{port05} R. Porter and D. Evans, ``Embedded Rayleigh-Bloch 
  surface waves along periodic rectangular arrays,''  Wave Motion 
  {\bf 43}, 29-50 (2005). 

\bibitem{shipman07} S. Shipman and D. Volkov, ``Guided modes in 
  periodic slabs: existence and nonexistence,''  
SIAM J. Appl. Math. {\bf 67}, 687--713 (2007). 

\bibitem{mari08} D. C. Marinica, A. G. Borisov, and 
  S. V. Shabanov, ``Bound states in the continuum in photonics,'' 
  \prl\ {\bf 100}, 183902 (2008).   



\bibitem{lee12} J. Lee, B. Zhen, S. L. Chua, W. Qiu, J. D. Joannopoulos, 
  M. Solja\v{c}i\'{c}, and O. Shapira, ``Observation and 
  differentiation of unique high-Q optical resonances near zero wave 
  vector in macroscopic photonic crystal slabs,'' \prl\ {\bf 109}, 
  067401 (2012). 

\bibitem{hsu13_2} C. W. Hsu, B. Zhen, J. Lee, S.-L. Chua, 
  S. G. Johnson, J. D. Joannopoulos, and M. Solja\v{c}i\'{c}, 
  ``Observation of trapped light within the radiation continuum,'' 
  Nature {\bf 499}, 188--191 (2013). 

\bibitem{bulg14b} E. N. Bulgakov and A. F. Sadreev, ``Bloch 
  bound states in the radiation continuum in a periodic array of 
  dielectric rods,''   \pra\ {\bf 90}, 053801 (2014). 

\bibitem{yang14} Y. Yang, C. Peng, Y. Liang, Z. Li, and S. Noda, ``Analytical 
perspective for bound states in the continuum in 
photonic crystal slabs,'' \prl\ {\bf 113}, 037401 (2014). 

\bibitem{zhen14} B. Zhen, C. W. Hsu, L. Lu, A. D. Stone, and M. 
Solja\v{c}i\v{c},  ``Topological nature of optical bound 
states in the continuum,'' 
\prl\ {\bf 113}, 257401 (2014). 

\bibitem{hu15} Z. Hu and Y. Y. Lu, ``Standing waves on two-dimensional 
  periodic dielectric waveguides,''  Journal of Optics {\bf 17}, 
  065601 (2015).  

\bibitem{gao16} X. Gao, C. W. Hsu, B. Zhen, X. Lin,
  J. D. Joannopoulos, M. Solja\v{c}i\'{c}, and H. Chen, ``Formation mechanism 
  of guided resonances and bound states in the continuum in photonic 
  crystal slabs,'' Sci. Rep. {\bf 6}, 31908 (2016). 

\bibitem{gan16} R. Gansch, S. Kalchmair,  P. Genevet, 
  T. Zederbauer, H. Detz,  A. M. Andrews, W. Schrenk, F. Capasso,  M. Lon\v{c}ar, and 
 G. Strasser, ``Measurement of bound states in the continuum by a 
 detector embedded in a photonic crystal,'' Light: Science \&
 Applications {\bf 5}, e16147 (2016). 

\bibitem{li16} L. Li and H. Yin, ``Bound states in the continuum in 
double layer structures,'' Sci. Rep. {\bf 6}, 26988 (2016). 

\bibitem{ni16} L. Ni, Z. Wang, C. Peng, and Z. Li, ``Tunable optical 
  bound states in the continuum beyond in-plane symmetry protection,'' 
\prb\ {\bf 94}, 245148  (2016). 

\bibitem{yuan17} L. Yuan and Y. Y. Lu, ``Propagating Bloch modes 
  above the lightline on a periodic array of cylinders,'' 
  J. Phys. B: Atomic, Mol. and Opt. Phys. {\bf 50}, 05LT01 (2017). 

\bibitem{bulg17pra} E. N. Bulgakov and D. N. Maksimov, 
``Bound states in the continuum and polarization singularities in 
periodic arrays of dielectric rods,'' 
\pra\ {\bf 96}, 063833 (2017). 

\bibitem{hu18} Z. Hu and Y. Y. Lu, ``Resonances and bound states in 
  the continuum on periodic arrays of slightly noncircular 
  cylinders,'' J. Phys. B: At. Mol. Opt. Phys.  {\bf 51}, 035402 (2018). 

\bibitem{doel18} H. M. Doeleman, F. Monticone, W. den Hollander, 
  A. Al\`{u}, and A. F. Koenderink, 
``Experimental observation of a polarization vortex at an optical 
bound state in the continuum,'' Nature Photonics {\bf 12}, 397--401 
(2018). 

\bibitem{zhang18} Y. Zhang, A. Chen, W. Liu, C. W. Hsu, B. Wang, F. Guan, 
  X. Liu, L. Shi, L. Lu, and J. Zi, 
  ``Observation of polarization vortices in momentum space,'' 
  \prl\  {\bf 120}, 186103 (2018). 


\bibitem{yuan17_4} L. Yuan and Y. Y. Lu,  ``Bound states in the 
  continuum on periodic structures: perturbation theory and 
  robustness,'' \ol\ {\bf 42}(21) 4490-4493 (2017). 

\bibitem{kosh18} K. Koshelev, S. Lepeshov, M. Liu, A. Bogdanov, and
  Y. Kibshar, ``Asymmetric metasurfaces with high-$Q$
    resonances governed by bound states in the contonuum,''
    \prl\, {\bf     121}, 193903 (2018).

\bibitem{lijun19pr} L.  Yuan and Y. Y.  Lu, ``Perturbation theories
  for symmetry-protected bound states in the continuum on
  two-dimensional periodic structures,'' 
arXiv preprint arXiv:1911.03612 (2019). 




\bibitem{lijun17pra} L.  Yuan and Y. Y. Lu, 
``Strong resonances on periodic arrays of cylinders and optical 
bistability with weak incident waves,'' 
\pra\ {\bf 95}, 023834 (2017). 

\bibitem{lijun18pra} L.  Yuan and  Y. Y.  Lu, 
  ``Bound states in the continuum on periodic structures surrounded by 
  strong resonances,'' 
\pra\ {\bf 97}, 043828 (2018). 

\bibitem{jin19} J. Jin, X. Yin, L. Ni, M. Soljacic, B. Zhen, and
  C. Peng, ``Topologically enable unltra-high-$Q$ guided resonances 
robust to out-of-plane scattering,''  Nature {\bf 574}, 501-504
(2019). 


\bibitem{kodi17} A. Kodigala, T. Lepetit, Q. Gu, B. Bahari,
  Y. Fainman, and B. Kant\'{e}, ``Lasing action from photonic bound 
  states in continuum,'' Nature {\bf 541}, 196-199 (2017).

\bibitem{romano19} S. Romano, A. Lamberti, M. Masullo, E. Penzo, 
  S. Cabrini, I. Rendina, and V. Mocella, 
  ``Optical biosensors based on photonic crystals
supporting bound states in the continuum,'' Materials {\bf 11}, 526 (2018). 

\bibitem{yesi19} F. Yesilkoy, E. R. Arvelo, Y. Jahani, M. Liu,
  A. Tittl, V. Cevher, Y. Kivshar, and H. Altug, ``Ultrasensitive
  hyperspectral imaging and biodetection enabled by dielectric
  metasurfaces,'' 
  Nature Photonics {\bf 13}, 390-396 (2019). 



\bibitem{lijun16pra} 
L. Yuan and Y. Y. Lu, 
``Diffraction of plane waves by a periodic array of nonlinear circular
cylinders,''
\pra\  {\bf 94}, 013852 (2016). 

\bibitem{lijun19siam}  L. Yuan and  Y. Y.  Lu, 
``Excitation of bound states in the continuum via second harmonic
generations,''   arXiv preprint arXiv:1908.00137 (2019). 


\bibitem{shipman05} S. P. Shipman and S. Venakides, ``Resonant 
  transmission near nonrobust periodic slab modes,''
\pre\  
{\bf 71},  026611 (2005). 


\bibitem{mocella15} V. Mocella and S. Romano, ``Giant field 
  enhancement in photonic lattices,'' \prb\ {\bf 92}, 155117 
  (2015). 

\bibitem{yoon15} J. W. Yoon, S. H. Song, and R. Magnusson, ``Critical 
  field enhancement of asymptotic optical bound states in the 
  continuum,'' Sci. Rep. {\bf 5}, 18301 (2015). 




\bibitem{fan02} S. Fan and J. D. Joannopoulos, ``Analysis of guided 
  resonances in photonic crystal slabs,'' 
\prb\ {\bf 65}, 235112 (2002). 


\bibitem{amgad19} A. Abdrabou and Y. Y. Lu, 
``Indirect link between resonant and guided modes on uniform and 
periodic slabs,'' 
\pra\ {\bf 99}, 063818 (2019). 

\bibitem{bao95}  G.  Bao, D. C.  Dobson, and  J. A.  Cox,
``Mathematical studies in rigorous grating theory,''
\josaa\ 
{\bf 12}(5), 1029-1042 (1995). 

\bibitem{huyuan19} Z.  Hu, L.  Yuan,  and Y. Y.  Lu, 
``Bound states with complex frequencies near the continuum on lossy
periodic structures,'' 
arXiv preprint arXiv:1910.02229 (2019). 

\end{thebibliography}
\end{document}